\documentclass{aa}
\usepackage[varg]{txfonts}

\usepackage{color}
\usepackage[english]{babel}
\usepackage[version=3]{mhchem}
\usepackage{setspace}

\title{Importance of tunneling in H-abstraction reactions by OH radicals}
\subtitle{The case of \ce{CH4 + OH} studied through isotope-substituted analogs}
\author{T. Lamberts\inst{1} \and G. Fedoseev\inst{2}\thanks{Current address: INAF–Osservatorio Astrofisico di Catania, via Santa Sofia 78, 95123 Catania, Italy} \and J. K\"astner\inst{1} \and S. Ioppolo\inst{3} \and H. Linnartz\inst{2}}

\institute{Institute for Theoretical Chemistry, University Stuttgart, Pfaffenwaldring 55, 70569 Stuttgart, Germany  \and Sackler Laboratory for Astrophysics, Leiden Observatory, Leiden University, PO Box 9513, 2300 RA Leiden, the Netherlands \and School of Physical Sciences, The Open University, Walton Hall, Milton Keynes, MK7 6AA, UK }

\date{Received XX/XX/XXXX / Accepted XX/XX/XXXX}

\abstract{We present a combined experimental and theoretical study focussing on the quantum tunneling of atoms in the reaction between \ce{CH4} and OH. The importance of this reaction pathway is derived by investigating isotope substituted analogs. 
Quantitative reaction rates needed for astrochemical models at low temperature are currently unavailable both in the solid state and in the gas phase. Here, we study tunneling effects upon hydrogen abstraction in \ce{CH4 + OH} by focusing on two reactions: \ce{CH4 + OD -> CH3 + HDO} and \ce{CD4 + OH -> CD3 + HDO}. The experimental study shows that the {solid-state} reaction rate $ R_\text{\ce{CH4 + OD}}$ is higher than $R_\text{\ce{CD4 + OH}} $ at 15~K. Experimental results are accompanied by calculations of the corresponding unimolecular and bimolecular reaction rate constants using instanton theory taking into account surface effects. From the work presented here, the unimolecular reactions are particularly interesting as these provide insight into reactions following a Langmuir-Hinshelwood process. The resulting ratio of the rate constants shows that the H abstraction ($ k_\text{\ce{CH4 + OD}}$) is approximately ten times faster than D-abstraction ($ k_\text{\ce{CD4 + OH}} $) at 65~K. We conclude that tunneling is involved at low temperatures in the abstraction reactions studied here. The unimolecular rate constants can be used by the modeling community as a first approach to describe OH-mediated abstraction reactions in the solid phase. For this reason we provide fits of our calculated rate constants that allow the inclusion of these reactions in models in a straightforward fashion. }

\begin{document}


 \maketitle

\section{Introduction}
 Radicals in ices are important players in solid-state interstellar chemistry \citep{Dishoeck:2013, Oberg:2016}. In particular, OH radicals seem to be abundant in ices \citep{Chang:2014, Lamberts:2014}; they play a role during solid-state water formation processes \citep{Cuppen:2010B,Oba:2012, Lamberts:2013, Lamberts:2016}, are involved in the formation of solid \ce{CO2} \citep{Oba:2010, Noble:2011, Ioppolo:2013}, and have been postulated to be crucial for the formation of complex organic molecules \citep{Garrod:2013, Acharyya:2015}. Abstraction reactions of the type \ce{OH + HC-R -> H2O + C-R} may form carbon-based radicals that can subsequently react with one another forming molecules that contain C--C bonds. Such abstraction reactions have recently been studied experimentally in the gas phase and show the importance of tunneling \citep{Shannon:2013, Caravan:2014}. 
In the solid state abstraction reactions along the hydrogenation line, \ce{CO - H2CO - CH3OH} was shown to provide extra pathways to form methylformate, ethylene glycol, and glycol aldehyde \citet{Chuang:2016}. % (Chuang et al. NRAS 455 (2016) 1702)
Although these gas-phase and solid-state reactions have been implemented in some astrochemical models, this is not yet common practice; for comparison, we refer the reader to \citet{Garrod:2013} and \citet{Acharyya:2015} to \citet{Vasyunin:2013} and \citet{Furuya:2014}, for example. Moreover, although a hydrogen transfer is involved, the role of tunneling at low temperature has not yet been investigated specifically. This is the main goal of the present work. 

{Tunneling can be defined as the quantum mechanical phenomenon that allows a particle to cross a barrier without having the energy required to surmount this barrier, that is, reaction routes that classically cannot occur at low temperatures become accessible. As a result of the mass dependence of tunneling processes, reactions with hydrogen, for instance, will be faster than with deuterium. }

{Although OH-mediated abstractions of hydrogen can occur with any hydrocarbon, the simplest example concerns \ce{CH4}.}
Interstellar methane is proposed to be a starting point of complex organic chemistry \citep{Oberg:2016}. It likely originates from carbon hydrogenation on the grain surface that occurs simultaneously with \ce{H2O} formation with abundances of $\sim$5\% \ce{CH4} with respect to \ce{H2O} \citep{Lacy:1991, Boogert:1998, Oberg:2008}.
\ce{H2O} ices are formed via hydrogenation of OH and therefore, OH will thus naturally be present in the ice \citep{Lamberts:2014}. Furthermore, weak FUV irradiation has been shown to result in photodissociation of water, which also yields OH \citep{Garrod:2013, Drozdovskaya:2016}. As mentioned above, hydroxyl radicals are known to react with CO and produce \ce{CO2} that has been abundantly seen and modeled in the water-rich layers of the ice. Since methane is present in the same layer, reactions of methane with OH are therefore expected to take place as well. {Note that these molecules have also recently been seen to co-exist in the coma of 67P/Churyumov-Gerasimenko \citep{Bockelee:2016}. Here,} we perform a proof-of-concept study on the importance of tunneling for the reaction \ce{CH4 + OH -> H2O + CH3}, both experimentally in the solid phase and through high-level calculations of reaction rate constants in the gas phase. The latter are performed in such a way that the results also allow us to draw conclusions on the solid state processes.

The experiments that are described in Section~\ref{ExpProc} make use of OH (or OD) produced in the solid phase. {Calculations performed by \citet{Arasa:2013, Meyer:2014} show that dissipation of excess energy takes place on a picosecond timescale. Therefore, thermalized reactions with the produced hydroxyl radicals occur as would be the case} in the interstellar medium (ISM). The gas-phase activation energy for hydrogen abstraction from \ce{CH4} is reported to be {$\sim$3160~K = 26.3 kJ/mol without zero-point energy (ZPE) correction \citep{Li:2015}}. For the reaction to proceed at cryogenic temperatures, tunneling needs to be efficient. To the best of our knowledge, rate constants below 200~K are not currently available, either in the gas phase or in the solid state \citep{Atkinson:2003}. Here, the role of tunneling is studied by comparing two sets of experiments, one probing H-abstraction -- \ce{CH4 + OD -> CH3 + HDO} -- and another probing D-abstraction -- \ce{CD4 + OH -> CD3 + HDO}.
These surface reactions have already been studied experimentally \citep{Wada:2006, Hodyss:2009, Weber:2009, Zins:2012}. {We extend on this previous work by quantitatively studying the isotope effect in the aforementioned abstraction reactions.}

The calculations that are described in Section~\ref{CompDet} incorporate tunneling through instanton theory, which allows reaction rate constants to be calculated at low temperatures. Also, previously published theoretical rate constants only reach down to 200~K, and, moreover, only bimolecular values were reported \citep{Wang:2012, Allen:2013}. In order to compare this with the present solid-state experiments, however, unimolecular rate constants are more relevant. These correspond to the decay rate of a pre-reactive complex of OH and \ce{CH4} on the surface, that is, from a configuration where the species have met while being thermalized, as is the case in the Langmuir-Hinshelwood (LH) mechanism. 

In Section~\ref{RD} the experimental results and theoretical calculations are presented. The final Section discusses the astrophysical consequences. {It is common for astrochemical models to make use of a rectangular barrier approximation in the description of tunneling; see \citet{Hasegawa:1992}, for example. We show here that kinetic isotope effects calculated while taking tunneling into account explicitly yield values that differ by several orders of magnitude with respect to instanton calculations.}
This work also ultimately shows that it is the combination of experiments, theory, and the method of implementation of results in models which is key to understanding the general behavior of tunneling in H-abstraction reactions by OH radicals.

\section{Methodology}

\subsection{Experimental procedure}\label{ExpProc}
The experiments have been performed using the ultra-high vacuum (UHV) setup SURFRESIDE. All relevant experiments are listed in Table~\ref{exps}. Experimental details for the setup configuration used here are available from \citet{Ioppolo:2008}. Briefly, a gold-coated copper substrate is placed in the center of a stainless steel UHV chamber ($P_{\text{base, main}}< 4\times 10^{-10}$ mbar) and kept at a temperature of 15~K using a He closed-cycle cryostat. For all experiments, a \ce{CH4}:\ce{O2} (\ce{CD4}:\ce{O2}) mixture is continuously deposited simultaneously with a D (H) beam. This is a so-called co-deposition experiment. Mixtures of \ce{CH4} or \ce{CD4} with \ce{O2} are deposited at an angle of 45$^{\circ}$ with a deposition rate of 0.45 Langmuir per minute, controlled by a precise all-metal leak valve, where 1 Langmuir corresponds to $\sim$1.3$\times10^{-6}$ mbar s$^{-1}$. Such an exposure rate corresponds to $\sim$2.7$\times 10^{12}$ molecules of \ce{CH4} or \ce{CD4} cm$^{-2}$ s$^{-1}$ and $\sim$6.7$\times 10^{11}$ molecules \ce{O2} cm$^{-2}$ s$^{-1}$. Gas mixture preparation is performed in a pre-pumped stainless steel dosing line ($P_\text{base, dosing line}<10^{-5}$ mbar) directly attached to the metal leak valve. A new mixture is prepared for each experiment. {The hydroxyl radicals are produced in the ice via hydrogenation or deuteration of \ce{O2}, {for example}, \ce{H + O2 -> HO2} (\ce{D + O2 -> DO2}) and \ce{H + HO2 -> 2 OH} (\ce{D + DO2 -> 2 OD}) \citep{Lamberts:2014}. This yields cold hydroxyl radicals via quick energy dissipation as explained above. Note that these hydroxyl radicals are thus formed during deposition, {that is}, methane molecules are in the direct vicinity of the formed OH radicals as a result of the 4:1 \ce{CH4}:\ce{O2} ratio. The H or D atoms needed for this} are generated in a thermal cracking source \citep{Tschersich:1998} facing the substrate. \ce{H2} or \ce{D2} molecules are dissociated after collisions with the hot walls ($T\approx2120$~K) of a tungsten capillary pipe, which in turn is surrounded by a heated tungsten filament. A nose-shaped quartz pipe along the beam path cools down the beam to roughly room temperature before it reaches the surface. 
{The dissociation fraction mentioned by \citet{Tschersich:2000} is $\sim$40\% at 2120~K for a pressure higher than that used here, inferring a higher dissociation ratio for the present experiments. However, this should be considered to be an upper limit, since H-H recombination may take place on the walls of the quartz pipe as well as on the ice surface. From \citet{Ioppolo:2010}, we can infer a \ce{H2}/H ratio of approximately 10-15\%. The H and D-fluxes used here are $\sim$6$\times 10^{13}$ and $\sim$3$\times 10^{13}$ atoms cm$^{-2}$ s$^{-1}$ \citep{Ioppolo:2010}. This ensures an overabundance of at least an order of magnitude of H (D) atoms with respect to \ce{CD4} (\ce{CH4}) and \ce{O2} on the surface.} %{The gases used are: \ce{CH4}, \ce{CD4}, \ce{O2}, \ce{H2}, and \ce{D2}}.

The ice composition is monitored using reflection absorption infrared spectroscopy (RAIRS) in the spectral range $700 - 4000$ cm$^{-1}$ ($14-2.5$ $\mu$m) with a spectral resolution of 0.5 cm$^{-1}$ and averaging 512 spectra,
using a Fourier transform infrared spectrometer. RAIR difference spectra are acquired with respect to a background spectrum of the bare gold substrate at 15~K.

\begin{table}[t]
 \centering\caption{List of performed co-deposition experiments (1 and 5) and control co-deposition experiments (2--4, 6,7).}\label{exps}
 \begin{tabular}{l@{\;\;}l@{}lll}
 \hline\hline
 No. & Ice mixture &  & Species & Duration  \\
 & & & & (min) \\
 \hline
 1   & \ce{CH4 : O2} & 4:1 & D \small (+ \ce{D2}) & 300 \\
 2   & \ce{CH4 : O2} & 4:1 & \ce{D2} & 300  \\
 3   & \ce{CH4}      &  & D \small (+ \ce{D2}) & 180  \\
 4   & \ce{CH4 : N2} & 1:4 & D \small (+ \ce{D2}) & 300  \\
 5   & \ce{CD4 : O2} & 4:1 & H \small (+ \ce{H2}) & 300 \\
 6   & \ce{CD4 : O2} & 4:1 & \ce{H2} & 196  \\
 7   & \ce{CD4} & & H \small (+ \ce{H2}) & 300  \\
 \hline
 \end{tabular}
\end{table}

\begin{figure}[t]
\centering
\resizebox{0.45\textwidth}{!}{\includegraphics{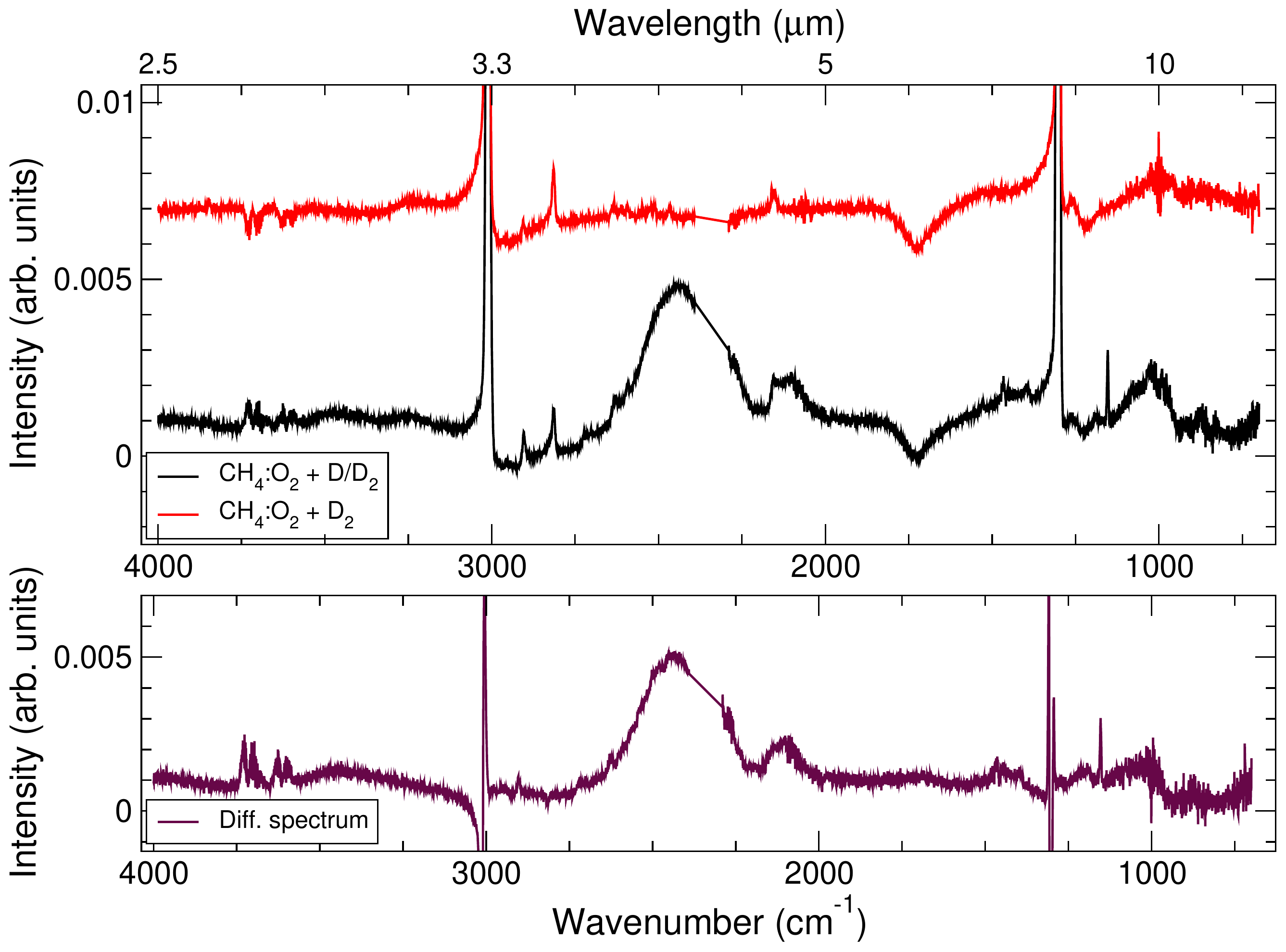}}
\caption{Full RAIR spectrum of exps. 1 (black) and 2 (red) and their difference curve (purple). See Table~\ref{assignment} for the assignment of the peaks. Note that the \ce{CO2} band has been removed for reasons of clarity.}
\label{CH4}
\end{figure}

\begin{figure}[t]
\centering
\resizebox{0.42\textwidth}{!}{\includegraphics{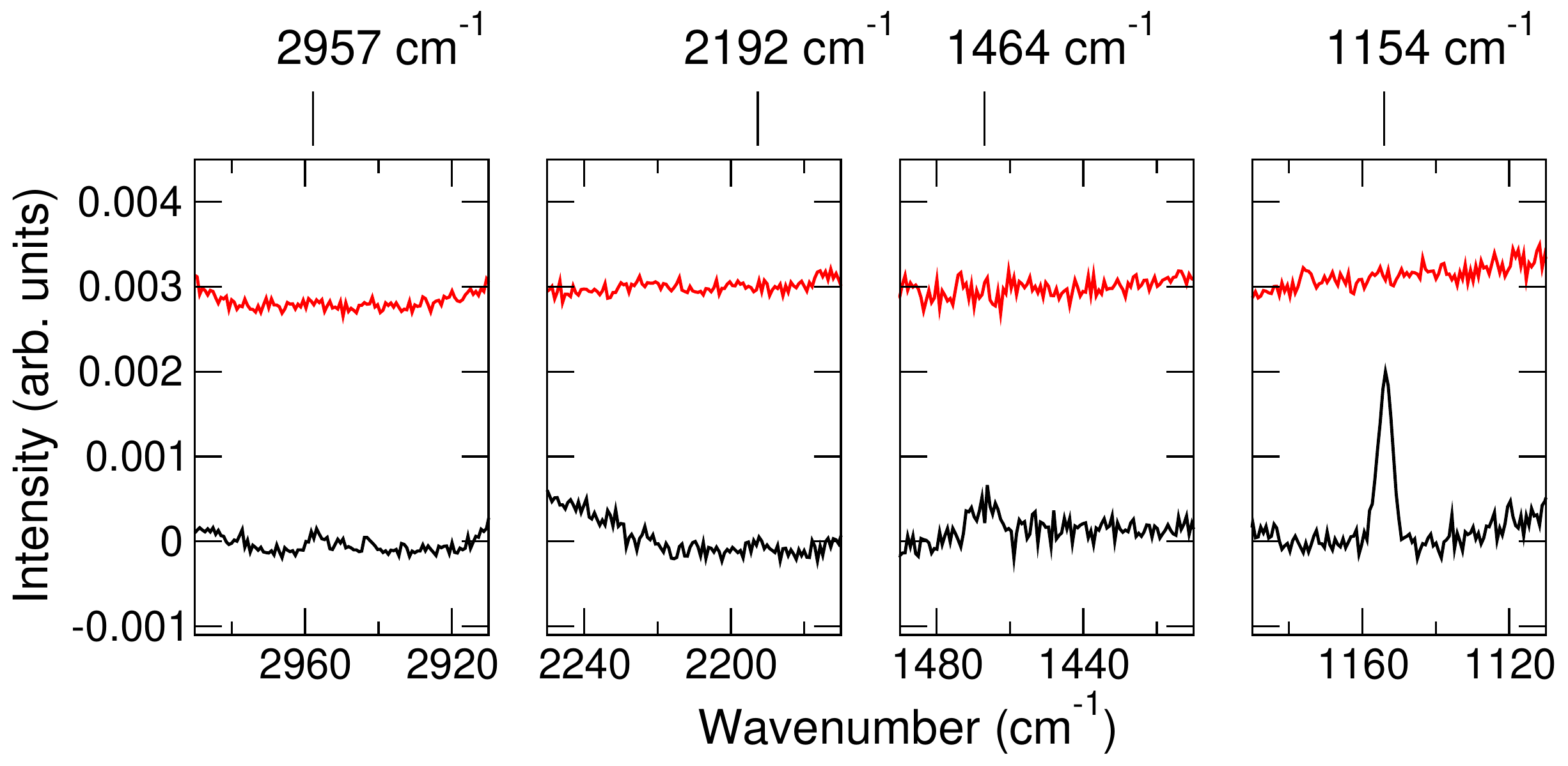}}
\caption{RAIR spectrum of exps. 1 (black) and 2 (red), zoomed-in on the regions of interest of known \ce{CH3D} IR transitions. See Table~\ref{assignment} for the assignment of the peaks.}
\label{CH4z}
\end{figure}

Straight baseline segments are subtracted from all acquired spectra, using five reference points: 699, 1140, 1170, 1830, and 4000 cm$^{-1}$. All spectra shown are normalized to a duration of 300 min (see Table~\ref{exps}) to allow direct comparison in the figures below. Furthermore, for all spectra, three data points are binned to allow for some smoothing. The reproducibility of the experiments is warranted by a triple execution and reproduction of exp. 1. The resulting data of only one of these measurements is shown in Figs.~\ref{CH4} and ~\ref{CH4z}.

{Control experiments have been chosen to exclude reactions not involving OD (OH) and to exclude the direct abstraction of H (D) from \ce{CH4} (\ce{CD4}) by D (H) atoms. Table~\ref{exps} provides an overview of the \ce{CH4 + OD} and \ce{CD4 + OH} experiments (exps. 1 and 5) and control experiments (exps. 2--4, 6, and 7).}
Control experiments 2 and 6 are directly related, since any products that are formed both in experiments 1 and 2 (or 5 and 6) are not caused by any reactions of the hydroxyl radical, because \ce{O2} and \ce{D2} (\ce{H2}) do not produce OD (OH). The control experiments 3, 4, and 7 aim to exclude the direct abstraction of H (D) from \ce{CH4} (\ce{CD4}) by D (H) atoms. That is, if no abstraction products are detected in control experiments 3, 4, and 7, any differences that are seen between experiments 1 and 5 are due to reactions involving the hydroxyl radical only.

\subsection{Computational details}\label{CompDet}
To obtain reaction rate constants that take tunneling into account explicitly, we use instanton theory \citep{Miller:1975, Callan:1977}. This is based on statistical Feynman path integrals that incorporate quantum tunneling effects, (see \citet{Kaestner:2014,Richardson:2016}). The instanton is the optimal tunneling path, which is usually different from the minimal energy path between stationary points (reactant, transition, and product states). Instanton theory is applicable when the temperature is low enough for the path to spread out, {or to put it differently, when tunneling dominates the reaction}. This is typically the case below the crossover temperature, $T_\text{C} = {\hbar \omega_b}/{2\pi k_\text{B}}$, where $\omega_b$ is the absolute value of the imaginary frequency at the transition state.

Instanton theory as implemented in DL-FIND \citep{Kaestner:2009, Rommel:2011} is used in this study on the potential energy surface (PES) of \citet{Li:2015} which is fitted to CCSD(T)-F12/AVTZ data. The Feynman paths of the instantons are discretized to 200 images. Instanton geometries are converged until the gradient is below $10^{-9}$ atomic units. 

The experimental data show that in the solid state, the reaction takes place between thermalized species (Section~\ref{ExpIR}). {These species have already approached one another through surface diffusion prior to reaction via the LH mechanism. The relevant rate constant calculated here describes the decay of this encounter complex. Therefore, we calculate unimolecular reaction rates excluding the probability of meeting, hence the unit of s$^{-1}$.} The vibrational adiabatic barrier for this process, that is, originating from the Van der Waals complex in the reactant channel, is 2575~K (21.4~kJ/mol) on the PES while it is slightly higher, 2670~K (22.2~kJ/mol), when calculated directly with CCSD (T)-F12/VTZ-F12 on the PES geometries. 

We model the surface by allowing for instant dissipation of energy through the constant temperature assumed in instanton calculations. Furthermore, the concentration of the reactants on a grain is higher than in the gas phase. We provide rate constants that are independent of the concentration, as this is dealt with  separately in astrochemical models. However, our structural model includes only CH4 and OH and not explicit surface atoms. On the surface, rotational motions are restricted and therefore the rotational partition function is kept constant, that is, frozen-out. However, the rotational symmetry factor \citep{Fernandez:2007} of 3 between the $C_{3v}$-symmetric Van der Waals complex and the transition state or instanton is taken into account. {Finally, another effect of the surface can be on the activation barrier of the reaction. This is neglected in our approach. In the experiment, the surface consists of \ce{CH4}, \ce{O2}, \ce{H2O2}, and \ce{H2O}. Both \ce{CH4} and \ce{O2} do not pose strong restrictions on the reactants in terms of interaction or steric hindrance, while the main contribution to the ice comes from \ce{CH4}. Therefore, the only assumption we make is that the minority species \ce{H2O} and \ce{H2O2} do not impact strongly on the reaction.}

Bimolecular rates related to the gas-phase mechanism are also calculated for comparison with gas-phase experiments, see Appendix~\ref{RRCg}. There, the full rotational partition function including the rotational symmetry factor of 12 is taken into account and CH$_4$ and OH reactant state structures are used that are calculated with ChemShell \citep{Metz:2014} on a CCSD(T)-F12/VTZ-F12 level using Molpro 2012 \citep{MOLPRO}.

{Although OH has two degenerate spin states in the $^2\Pi$ ground state, when reacting with closed-shell molecules, both states are reactive and no additional correction factors for the rate constants are needed \citep{Graff:1990, Fu:2010}.}

\begin{table}[b]
 \centering
 \caption{Peaks identified in the difference spectrum between exps. 1 and 2, depicted in Fig.~\ref{CH4}.}\label{assignment}
 \begin{tabular}{llllll}
  \hline\hline
$\tilde{\nu}$  & \multicolumn{3}{l}{Specification} & Species & Ref.  \\
 (cm$^{-1}$) & $^{a}$ & $^{b}$ & $^{c}$ & & \\
  \hline
3434    & w & b &   &   \ce{HDO} & 1 \\ 
3015    & s & n &   &   \ce{CH4} & 2 \\ 
2957    & w &   &   &   \ce{CH3D}& 3 \\ 
2904    & m & n &   &   \ce{CH4} & 2 \\ 
2817    & m & n &   &   \ce{CH4} & 2 \\ 
2718    & w & b & u &            &   \\ 
2620    & w &   & p &   \ce{CH3D}& 3 \\ 
2440    & s & b &   &   \ce{D2O2}& 1 \\ 
2192    & w &   &   &   \ce{CH3D}& 3 \\ 
2115    & m & b &   &   \ce{D2O2}& 1 \\ 
1530    & w & b & p & \ce{CH3D} & 3 \\
1464    & w & n &   & \ce{CH3D} & 3 \\
        & w & b & p & \ce{HDO}  & 1 \\
1400    & w & b & u &           & \\
1304    & s & n &   & \ce{CH4}  & 2 \\
1195    & w & b & u &           & \\
1154    & m & n &   & \ce{CH3D} & 3 \\
1020    & s & b &   & \ce{D2O2} & 1 \\
870     & s & b &   & \ce{D2O2} & 4 \\
\hline
 \multicolumn{6}{l}{$^{a}$ w = weak, m = medium, s = strong} \\
 \multicolumn{6}{l}{$^{b}$ b = broad, n = narrow}\\
 \multicolumn{6}{l}{$^{c}$ u = unidentified, p = possible} \\
 \end{tabular}
  \tablebib{(1) \citet{Oba:2014}; (2) \citet{Edling:1987}; (3) \citet{Momose:2004}; (4) \citet{Giguere:1975} }
 \end{table}

In order to provide input for astrochemical modelers, we fit a rate expression, Eqn.~\ref{Zheng} \citep{Zheng:2010}, to the calculated rate constants; 
\begin{equation}
 k = \alpha \left(\frac{T}{300}\right)^\beta \exp\left({-\frac{\gamma (T + T_0)}{(T^2 +T_0^2)}}\right) \label{Zheng}
.\end{equation}
This expression is more suitable for describing tunneling reactions than the commonly used Kooij expression. {The parameters $\alpha$, $\beta$, $\gamma$, and $T_0$ are all fitting parameters, where $\alpha$ has the units of the rate constant, $\beta$ regulates the low-temperature behavior, and $\gamma$ and $T_0$ can be related to the activation energy of the reaction. $\beta$ is set to 1 since this results in the correct low-temperature behavior.} Instanton rate calculations are used for the fits below the crossover temperature $T_\text{C}$, while rate constants obtained from transition state theory - including quantized vibrations and a symmetric Eckart model for the barrier - are used above $T_\text{C}$ where tunneling plays a minor role. 

As a final remark, it is important to keep the difference between the reaction rate, $R$, and the rate constant, $k$, in mind. {Experimentally, only effective rates or the ratio of effective rates can be determined. This is because, experimentally, one can only detect the change in the production of a final product, meaning it is not possible to separate the diffusion and reaction processes. Theoretically, however, a rate constant that is related exclusively to the decay of the encounter complex can be calculated, {that is}, after the diffusion has already taken place.} Therefore, the kinetic isotope effect (KIE) for H-abstraction vs. D-abstraction derived from the effective rate and derived from the rate constants do not necessarily need to be identical.


\section{Results \& discussion}\label{RD}

\subsection{Experimental -- infrared spectra}\label{ExpIR}

\begin{figure}[t]
\centering
\resizebox{0.45\textwidth}{!}{\includegraphics{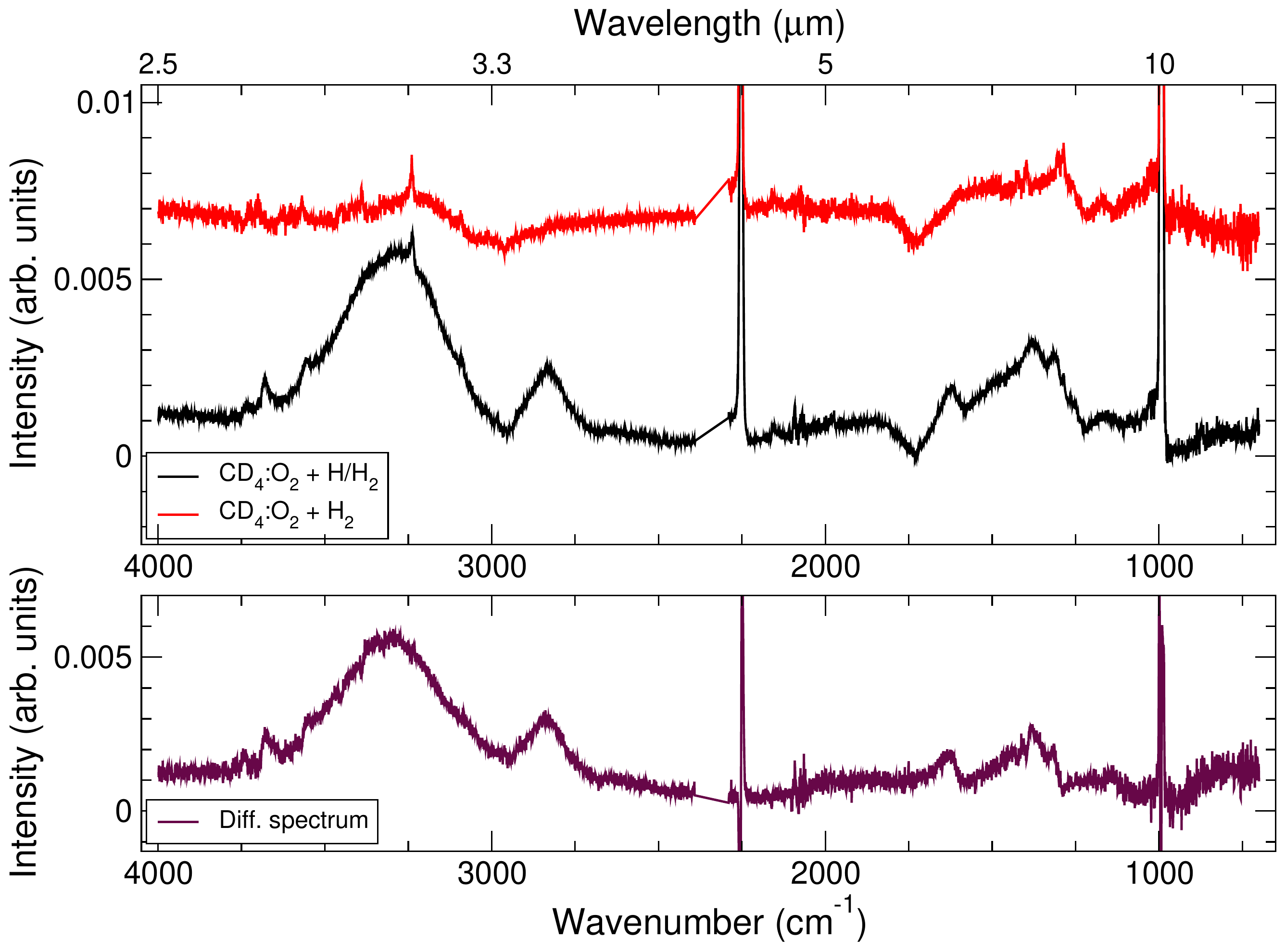}}
\caption{Full RAIR spectrum of exps. 5 (black) and 6 (red) and their difference curve (purple). Note that the \ce{CO2} band has been removed for reasons of clarity.}
\label{CD4}
\end{figure}

\begin{figure}[t]
\centering
\resizebox{0.42\textwidth}{!}{\includegraphics{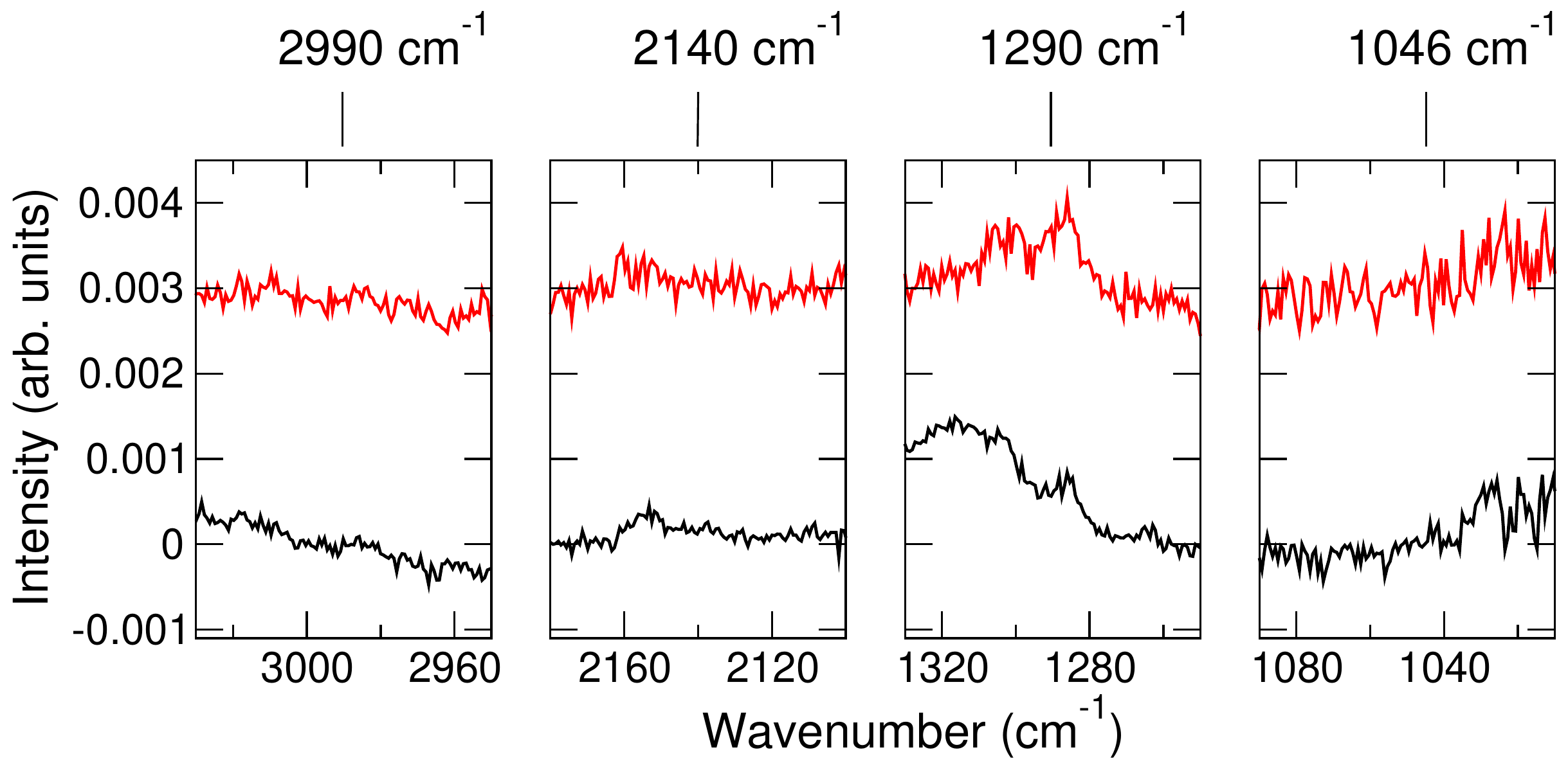}}
\caption{RAIR spectrum of exps. 5 (black) and 6 (red), zoomed-in on the regions of interest of known \ce{CD3H} IR transitions.}
\label{CD4z}
\end{figure}

Figures~\ref{CH4} and~\ref{CH4z} depict the final RAIR spectra acquired for exps. 1 and 2 (upper panels). The difference spectrum between exps. 1 and 2 is also given (lower panels), either showing the full spectral window (Fig.~\ref{CH4}) or zoomed-in on the regions of interest of \ce{CH3D} (Fig.~\ref{CH4z}). The same holds for exps. 5 and 6 (Fig.~\ref{CD4}), and zoomed-in on the regions of interest of \ce{CD3H} (Fig.~\ref{CD4z}).
In Table~\ref{assignment}, the peak positions that can be determined in the difference spectrum between exps. 1 and 2 are listed with their corresponding assignment and reference.

{The main outcome of the surface experiments is that at our detection level, abstraction from methane is only found to take place during exp. 1.} The reactants available there, \ce{CH4}, \ce{O2}, and \ce{D}, can undergo several reactions. The relevant ones are listed below. %\vspace{-10pt}
\begin{align}
&\ce{O2 + D} \rightarrow \ce{DO2} \tag{R1}\label{R1}\\  
&\ce{DO2 + D} \rightarrow 2\;\ce{OD} \tag{R2}\label{R2}\\
&2\;\ce{OD} \rightarrow \ce{D2O2} \tag{R3}\label{R3}\\
&\ce{D + D2O2} \rightarrow \ce{D2O + OD} \tag{R4}\label{R4}\\
&\ce{CH4 + OD} \rightarrow \ce{CH3 + HDO} \;\; \tag{R5}\label{R5} \\
&\ce{CH3 + D} \rightarrow \ce{CH3D} \;\; \tag{R6}\label{R6} \\
&\ce{CH3 + OD} \rightarrow \ce{CH3OD} \;\; \tag{R7}\label{R7} \\
&\ce{CH3 + CH3} \rightarrow \ce{C2H6} \;\; \tag{R8}\label{R8}
\end{align}
The equivalent reactions, but with deuterium substituted for hydrogen (starting from \ref{R1}), are applicable for experiment 5. {Note that the reaction \ce{CH4 + D -> CH3 + HD} has been excluded from the list of possible reactions. This is explained below.}


We clearly identify \ce{D2O2} and \ce{CH3D} infrared features and a weak stretching mode of \ce{HDO} in Fig.~\ref{CH4}; see Table~\ref{assignment} for the various peak positions. The bending mode of \ce{HDO} at 1475 cm$^{-1}$ is not clearly visible as a result of the signal-to-noise (S/N) level. The bending mode is at least four times weaker than the stretching mode, which is not detectable for the S/N ratio achieved here \citep{Oba:2014}. This means that the produced \ce{OD} radicals either form \ce{D2O2} or abstract an H-atom from \ce{CH4}. The availability of D atoms on the surface results in the barrierless deuteration of a \ce{CH3} radical. Since the D atoms are abundant and diffuse quickly, other radical-radical reactions are prevented. The \ce{CH3OD} or \ce{C2H6} yields lie below the sensitivity of our RAIRS technique, that is, reactions~\ref{R7} and~\ref{R8} are not relevant here. Note that the abstraction reaction between \ce{HO2} and \ce{CH4} has a barrier of $>10000$~K \citep{Aguilera:2008}, and we therefore exclude any \ce{CH3} radical production via this route. 

Furthermore, none of the control experiments (numbers 2-4) resulted in the detection of \ce{CH3D}. This means that the \ce{CH3D} detected in experiment 1 is not the cause of contamination (follows from experiment 2) nor the result of H-abstraction from \ce{CH4} by D atoms directly (follows from control experiments 3 and 4). {In other words, in our experimental setup, the reaction between \ce{CH4} and the D atom does not proceed within our experimental sensitivity. Since the same setup is used for
experiments 1, 3, and 4, essentially for identical conditions, we conclude that the reaction \ce{CH4 + OD} is more efficient and the only one responsible for the formation of the observed \ce{CH3D}.} {This result can be rationalized by realizing that the reaction \ce{CH4 + H -> CH3 + H2} has a high barrier of 6940-7545~K \citep{Nyman:2007, Corchado:2009}, making it highly unlikely to be competitive at the low temperatures considered here, even if it were to occur via tunneling. Indeed, the (bimolecular) rate constants calculated on several PES at 300~K are six orders of magnitude lower than the bimolecular rate constants calculated here for \ce{CH4 + OH} (compare \citealt{Corchado:2009}, \citealt{Espinosa-Garcia:2009}, and \citealt{Li:2013} with values mentioned in Appendix~\ref{RRCg}).}

The next question is whether or not the isotope-substituted reaction \ref{R5}, \ce{CD4 + OH -> CD3 + HDO,} leads to D-abstraction by OH radicals. In this case, the mass of the atom that is transferred is twice as high and if tunneling is important, the reaction rate should drop significantly. In Fig.~\ref{CD4}, and especially from the zoomed-in spectra in Fig.~\ref{CD4z}, it is clear that no detectable amounts of \ce{CD3H} are produced during experiment 5. Since OH radicals are present, \ce{H2O2} is formed and can be seen at its respective peak positions, around 3315, 2835, and 1385 cm$^{-1}$ \citep{Oba:2014}. Furthermore, the OH radicals that are produced are not being used for reactions with \ce{CD4} and therefore are available for other reactions, leading to the formation and detection of the \ce{H2O} bending mode at 1630 cm$^{-1}$, while the stretching mode overlaps with that of \ce{H2O2}. {\ce{H2O} can also be formed via \ce{H + H2O2 -> H2O + OH} in contrast to the formation of \ce{D2O} via \ce{D + D2O2 -> D2O + OD} as a result of the kinetic isotope effect \citep{Lamberts:2016b}}. 
The control experiments 6 and 7 do not result in the detection of any \ce{CD3H} either. 

Since neither experiment 1 nor 5 results in breaking of the bond between the carbon atom and hydrogen or deuterium atom, we can be certain that the OD and OH radicals that are formed in the ice thermalize prior to attempting to react. 
We propose that the origin of the difference between experiments 1 and 5 is the tunneling of the H or D atom that is being transferred from the carbon atom to the oxygen atom. {Experimentally, it is not trivial to quantify the ratio between the effective rates, but we can state that $R_\text{\ce{CD4 + OH}}$ is smaller than $R_\text{\ce{CH4 + OD}}$. A further quantification is the aim of the following section, using theoretical calculations of rate constants.}

 \begin{figure}[t]
\centering
{\includegraphics[height=3cm]{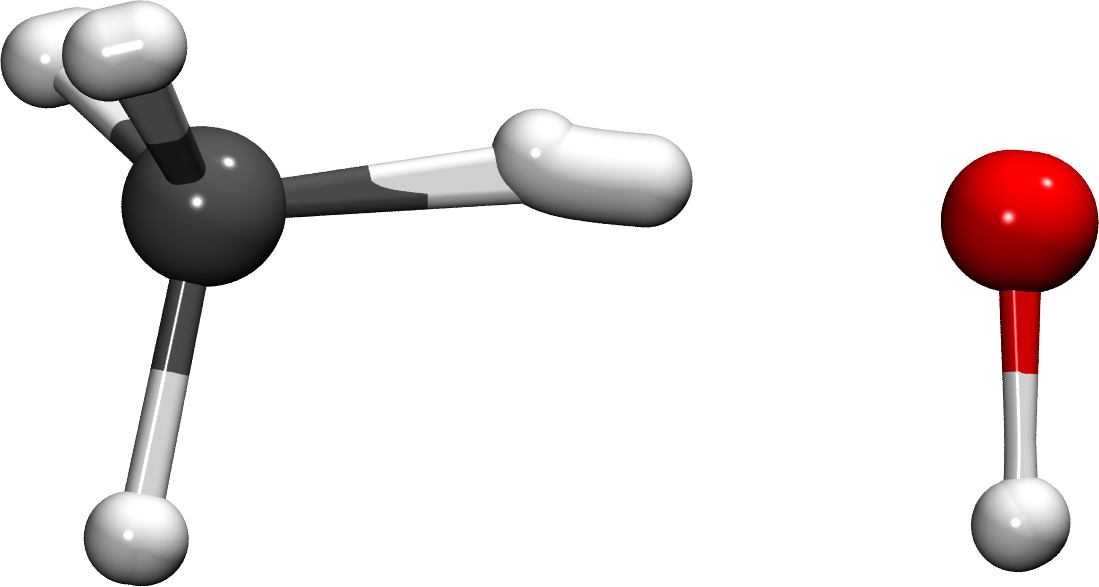}}
\caption{ Instanton path for \ce{CH4 + OH} at 200~K.  }
\label{highT}
\end{figure}
\begin{figure}[t]
\centering
{\includegraphics[height=3cm]{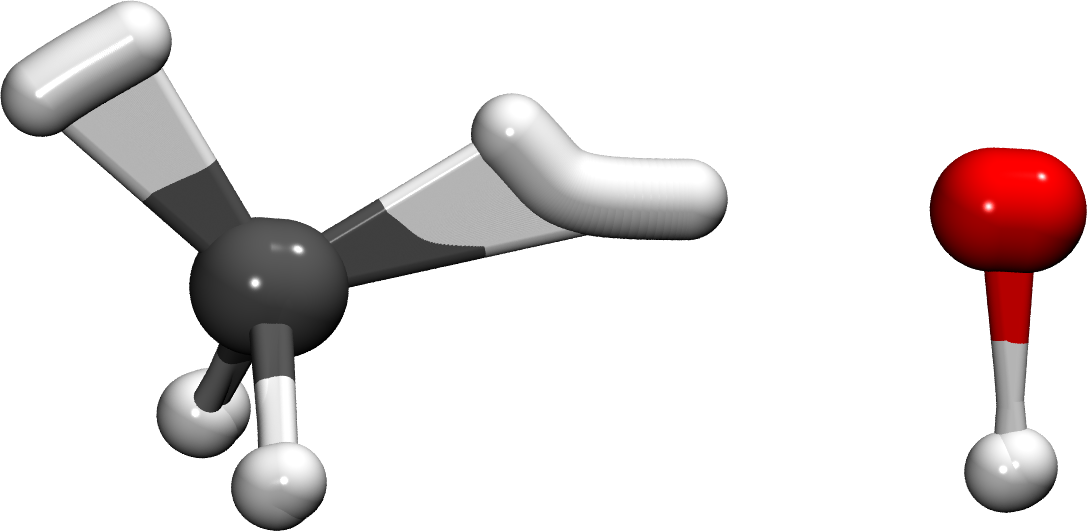}}
\caption{ Instanton path for \ce{CH4 + OH} at 100~K.  }
\label{lowT}
\end{figure}

\subsection{Theoretical -- LH unimolecular rate constants}

Rate constants are calculated using instanton theory. 
An instanton path essentially shows the delocalization of the atoms involved in the reaction. This deviates from the classical picture of overcoming a barrier and visualizes the tunneling through a barrier. More delocalization is seen at lower temperatures, where tunneling indeed plays a larger role.
In Fig.~\ref{highT} the instanton path is depicted for the reaction between \ce{CH4} and \ce{OH} at 200~K. This geometry corresponds to the geometry of the optimized transition state where the O--H bond is in an eclipsed configuration with one C--H bond. At low temperature, that geometry becomes unstable and the instanton switches to a staggered position. Fig.~\ref{lowT} gives the corresponding geometry at 100~K. This instability leads to a temperature region where instanton rates become unreliable \citep{Meisner:2011}. In the current paper, we only give rate constants that do not lie within the switching regime. {Therefore, not all rate constants can be calculated at exactly the same temperatures, as is visible in Table~\ref{LHrate}.} {Furthermore, rates can only be calculated down to a certain temperature (here, 65~K), because of inaccuracies in the PES around the Van der Waals complexes that become important at low temperatures. The unimolecular reaction rate is namely calculated as the decay of the Van der Waals encouter or pre-reactive complex.}
The activation energies including ZPE correction and starting from a pre-reactive minimum for the reactions \ce{CH4 + OH}, \ce{CD4 + OH} and \ce{CH4 + OD} are 2575, 2915, and 2605~K (21.1, 24.2, and 21.7 kJ/mol), respectively. 

\begin{figure}[t]
\centering
\resizebox{0.49\textwidth}{!}{\includegraphics{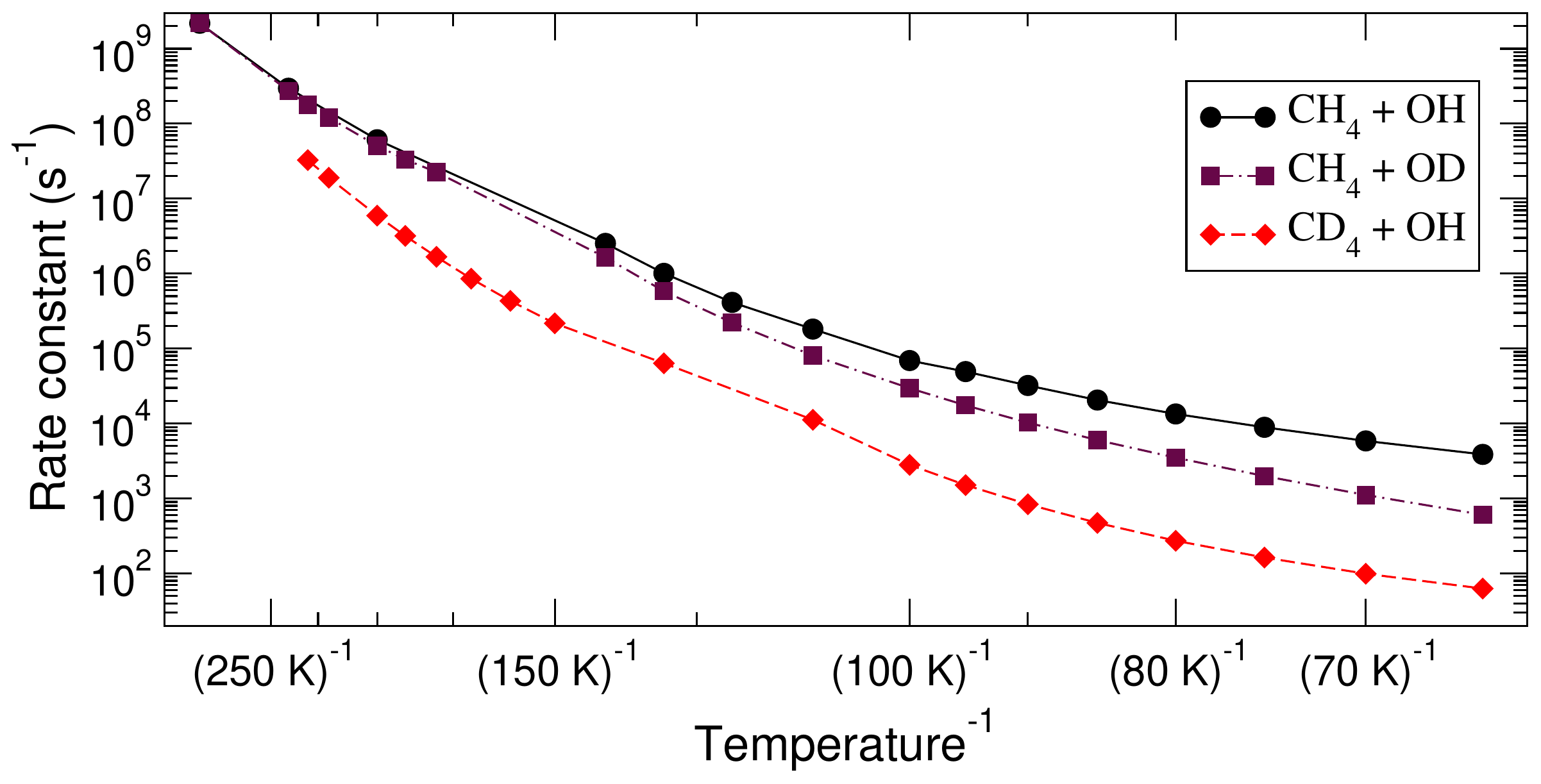}}
\caption{Arrhenius plot of the unimolecular reaction rate constants for three isotopological analogs of reaction \ref{R5}: \ce{CH4 + OH} (black, solid), \ce{CH4 + OD} (purple, dashed-dotted), and \ce{CD4 + OH} (red, dashed). }
\label{unimol}
\end{figure}

The unimolecular reaction rate is of interest for surface reactivity, since most interstellar-relevant reactions will take place between two reactants that have adsorbed on a grain prior to the reaction. Such reactants are thermalized and can only react after finding each other. As long as they are close together, they may attempt to react several times. In Fig.~\ref{unimol}, the corresponding calculated rate constants are depicted. {In general, rate constant values decrease with temperature.} The exact values are given in Table~\ref{LHrate}, while {Table~\ref{fits}} lists the parameters fitted to Eqn.~\ref{Zheng}. {A kinetic isotope effect ${k_\text{\ce{CH4 + OD}}}/{k_\text{\ce{CD4 + OH}}} $ of approximately ten is obtained between 200~and 65~K. Therefore, a similar value for the KIE at even lower temperatures seems to be a reasonable estimation. This is consistent with the conclusion derived in the previous paragraph for the experiments; the reaction \ce{CH4 + OD} is found to take place whereas for \ce{CD4 + OH} no reaction products are found, {that is}, the first reaction is definitely faster than the second.}

\section{Astrophysical Implications}

For the calculation of the reaction rate constants, we have focused here on the reaction between OH radicals and \ce{CH4} molecules because it provides a simple system that can be potentially extended to other H-abstraction reactions involving OH radicals and larger hydrocarbons as was proposed by \citet{Garrod:2013} and \citet{Acharyya:2015}. As mentioned before, OH radicals are present in the \ce{H2O} -rich layers of the ice as a result of non-energetic hydrogenation reactions and weak FUV irradiation that photodissociates water. Therefore, it is expected that they are available for reactions \citep{Garrod:2013}. The abundances of solid \ce{CH4} can reach levels of several percent with respect to water ice in a variety of environments, such as, low- and high-mass young stellar objects, quiescent clouds, cold cores, and cometary ices (\citealt{Oberg:2008, Boogert:2015, LeRoy:2015} and references therein). Moreover, methane is mostly formed in a water-rich environment \citep{Boogert:2015}. Therefore, the reaction \ce{CH4 + OH} is likely to occur in space at low temperatures. An important product of the aforementioned H-abstraction reaction is the methyl radical. In the ISM, solid \ce{CH3} is an intermediate reaction product in the formation of methane. It can also be formed through UV-, electrons- and ions-induced dissociation reactions from ice mixtures containing methane \citep{Milligan:1967, Bohn:1994, Palumbo:2004, Hodyss:2011, Kim:2012, Wu:2012, Wu:2013, Materese:2014, Materese:2015, Lo:2015, Bossa:2016, Chin:2016}. The solid \ce{CH4 + OH} abstraction reaction investigated here, however, can be an important channel for the formation of methyl radicals under interstellar analog conditions. It allows \ce{CH3} production without the need for photodissociation and strong energetic processing of the ice.

Recent disk and protostellar models showed that several molecules with a methyl group (\ce{R-CH3}) are present in ices \citep{Vasyunin:2013, Furuya:2014, Walsh:2014, Taquet:2015, Drozdovskaya:2016}. Moreover, one of the outcomes of the Rosetta mission is the discovery of heavy organic matter on the comet 67P/Churyumov-Gerasimenko \citep{Goesmann:2015, Fray:2016}. Also in this case, several species contain a methyl group. Therefore, surface reactions involving the \ce{CH3} functional group are assumed to play an important role in enhancing the molecular complexity in space. Large hydrocarbon radicals can be created by the H-abstraction from any larger molecule of the form \ce{R-CH3} , which is then available for subsequent reactions. 

{The fits to Eqn.~\ref{Zheng} can be used by the astrochemical modeling community as an initial approach to describing H-abstraction reactions from methane in the solid phase via the Langmuir-Hinshelwood mechanism. These fits are definitely valid down to temperatures of 65~K and we can recommend extrapolation down to $\sim$35~K. Although the desorption temperature of pure methane is relatively low, when it is trapped in the water ices, it is expected to desorb only when water does, that is, around 100~K \citep{Collings:2004}. In order to assure a smooth implementation of such rate constants in models, it is crucial to realize that the expressions typically used in rate equation models to describe tunneling involve the rectangular barrier approximation. Then, reaction rates are proportional to a probability $P = \exp(\frac{2a}{\hbar}\sqrt{2\mu E})$.} With a typical barrier width $a$ of 1 \AA, the activation energies mentioned above, and masses $\mu$ of 1 or 2 amu depending on H- or D-abstraction, as well as the ratio between the probabilities, $P_\text{\ce{CH4 + OD}}$ and $P_\text{\ce{CD4 + OH}}$, and thus between the rates, would be approximately $5\times10^3$. Here, {we show with the use of instanton calculations, that the KIE is approximately ten at low temperature, which is orders of magnitude lower than the value generally used in rate equation models.} {This implies that current models may need to improve on their implementation of H- vs. D-abstraction reactions. Although here we show the effect for one reaction only, this conclusion also holds for other astrochemically relevant reactions, such as \ce{H + H2O2 -> H2O + OH} \citep{Lamberts:2016b}.}

\begin{table}[t]
 \centering
 \caption{Unimolecular reaction rate constants.}\label{LHrate}
 \begin{tabular}{llll}
  \hline\hline
        & {\ce{CH4 + OH}} &   {\ce{CD4 + OH}} &   {\ce{CH4 + OD}} \\
        & {$T_\text{C} = 323.5$ K } & {$T_\text{C} = 240.2 $ K } & {$T_\text{C} = 323.1$ K } \\
        \hline
$T$ & $k$ & $k$ & $k$ \\
(K) & (s$^{-1}$) & (s$^{-1}$) & (s$^{-1}$) \\
\hline
300 & 2.18 $\times$ $10^9$      &                       & 2.19  $\times$ $10^9$          \\
240 & 2.97 $\times$ $10^8$      &                       & 2.73  $\times$ $10^8$          \\
230 &                           & 3.26  $\times$ $10^7$ & 1.78  $\times$ $10^8$  \\
220 &                           & 1.90  $\times$ $10^7$ & 1.20  $\times$ $10^8$          \\
200 & 6.10 $\times$ $10^7$      & 5.93  $\times$ $10^6$ & 5.07  $\times$ $10^7$          \\
190 &                           & 3.19  $\times$ $10^6$ & 3.32  $\times$ $10^7$          \\
180 &                           & 1.67  $\times$ $10^6$ & 2.25  $\times$ $10^7$          \\  
170 &                           & 8.57  $\times$ $10^5$ &                               \\
160 &                           & 4.32  $\times$ $10^5$ &                               \\
150 &                           & 2.17  $\times$ $10^5$ &                               \\
140 & 2.53 $\times$ $10^6$      &                       & 1.64  $\times$ $10^6$           \\
130 & 1.01 $\times$ $10^6$      & 6.36  $\times$ $10^4$ & 5.83  $\times$ $10^5$          \\
120 & 4.14 $\times$ $10^5$      &                       & 2.22  $\times$ $10^5$         \\
110 & 1.82 $\times$ $10^5$      & 1.12  $\times$ $10^4$ & 8.07  $\times$ $10^4$          \\
100 & 6.94 $\times$ $10^4$      & 2.81  $\times$ $10^3$ & 2.96  $\times$ $10^4$           \\
95 & 4.94 $\times$ $10^4$       & 1.51  $\times$ $10^3$ & 1.75  $\times$ $10^4$    \\
90 & 3.22 $\times$ $10^4$       & 8.37  $\times$ $10^2$ & 1.03  $\times$ $10^4$   \\
85 & 2.06 $\times$ $10^4$       & 4.74  $\times$ $10^2$ & 6.01  $\times$ $10^3$   \\
80 & 1.34 $\times$ $10^4$       & 2.73  $\times$ $10^2$ & 3.51  $\times$ $10^3$  \\
75 & 8.91 $\times$ $10^3$       & 1.63  $\times$ $10^2$ & 1.99  $\times$ $10^3$   \\
70 & 5.86 $\times$ $10^3$       & 9.93  $\times$ $10^1$ & 1.11  $\times$ $10^3$   \\
65 & 3.88 $\times$ $10^3$       & 6.29  $\times$ $10^1$ & 6.08  $\times$ $10^2$   \\
%60 & 2.60 $\times$ $10^3$      & 4.17  $\times$ $10^1$ & 3.24  $\times$ $10^2$    \\
  \hline                     
 \end{tabular}               
 \end{table} 
 
\begin{table}
 \centering
 \caption{Parameters fitted down to 65~K to describe the abstraction reactions \ce{CH4 + OH}, \ce{CD4 + OH}, and \ce{CH4 + OD}, respectively. }\label{fits}
 \begin{tabular}{llll}
 \hline\hline
 Parameter & \ce{CH4 + OH} & \ce{CD4 + OH} & \ce{CH4 + OD} \\
 \hline
 $\alpha$ (s$^{-1}$)    & 1.24 $\times$ $10^{11}$       & 5.86  $\times$ $10^{10}$       & 1.45  $\times$ $10^{11}$ \\
 $\beta$                & 1 & 1 & 1 \\
 $\gamma$ (K)           & 1201 & 1468 & 1252 \\
 $T_0$ (K)              & 83.1 & 85.0 & 73.9 \\
 \hline
 \end{tabular}
\end{table}

\section{Conclusions}\label{sect:AC}

With this combined laboratory-theoretical study, we provide a proof-of-principle for the importance of incorporating tunneling in the description of OH-mediated hydrogen abstraction. First, through solid phase experiments, we find that the surface reaction \ce{CH4 + OD} indeed proceeds faster than \ce{CD4 + OH} at 15~K as expected for tunneled reactions. Second, instanton calculations of unimolecular rate constants quantify the processes and give a kinetic isotope effect of approximately ten at 65~K. Such unimolecular rate constants relate to the thermalized Langmuir-Hinshelwood process, the main mechanism used to explain the non-energetic surface formation of species on and in interstellar ices. For the first time, rate constants at temperatures down to 65~K for the \ce{CH4 + OH} reaction are provided with accompanying fits to allow easy implementation into astrochemical models. {The calculated rate constants also show that the rectangular barrier approximation currently used in models inaccurately estimates kinetic isotope effects for the title reaction by approximately two orders of magnitude.}

\begin{acknowledgements}
We kindly thank Jun Li and Hua Guo for providing the potential energy surface.
Astrochemistry in Leiden in general was supported by the European Community’s Seventh Framework Programme (FP7/2007–2013) under grant agreement no. 238258, the Netherlands Research School for Astronomy (NOVA) and from the Netherlands Organization for Scientific Research (NWO) within the framework of the Dutch Astrochemistry Network. SI acknowledges the Royal Society for financial support. TL was supported by the Dutch Astrochemistry Network financed by NWO. TL and JK were financially supported by the European Union's Horizon 2020 research and innovation programme (grant agreement No. 646717, TUNNELCHEM).
\end{acknowledgements}

\singlespacing

\bibliographystyle{aa}
\bibliography{biblioCH4OH}

\begin{appendix}

\section{Gas-phase bimolecular reaction rate constants}\label{RRCg}

In the gas phase, rotational motion is not restricted, and moreover the symmetry of the reactant and transition states needs to be taken into account in the calculation of the rate constants \citep{Fernandez:2007}. The symmetry factor used here is 12, resulting from the pointgroups $T_d$ for the methane molecule, $C_{\infty V}$ for OH, and $C_s$ for the eclipsed and staggered instanton configurations. Furthermore, here, bimolecular rate constants are applicable that include the chance of meeting, hence the different units compared to unimolecular rate constants.

Table~\ref{tabapp} shows the calculated bimolecular reaction rate constants in the range 300--70~K.
The highest temperatures calculated here can be compared to the values at the lower end of the range recommended by \citet{Atkinson:2003}, or, more specifically, to the gas-phase experimental data points around 200~K provided by \citet{Gierczak:1997}. {We choose the lowest temperature possible, since the influence of tunneling will then be more prominent.} This shows that the maximum deviation between our instanton rate constants and their experiments is one order of magnitude. 

This overestimation has two origins, firstly the harmonic approximation in instanton theory is known to lead to an overestimation of the rate constants at temperatures close to the crossover temperature \citep{Goumans:2011}. At the low temperatures that we are interested in for surface reactivity, however, the harmonic approximation is valid and the part of the overestimation that stems from this thus vanishes. Furthermore, the lower barrier on the PES with respect to CCSD(T)-F12 calculations (see Section~\ref{CompDet}) leads to a small overestimation of the rate constant as well. 

The ratio between H- and D-abstraction by the \ce{OH} radical, ${k_\text{\ce{CH4 + OH}}}/{k_\text{\ce{CD4 + OH}}}$, is the same for our calculations and for the gas-phase experimental values \citep{Gierczak:1997}, namely $\sim$12.5. Additionally, the ratio ${k_\text{\ce{CH4 + OH}}}/{k_\text{\ce{CH4 + OD}}}$ is also similar, $\sim$0.73 in both cases. These ratios are calculated in the range where the calculations overlap with the gas-phase experiments,that is, around 230~K.
This agreement with bimolecular gas-phase experimental data provides a good basis for quantifying unimolecular surface processes.

\begin{table}[h]
 \centering
 \caption{Bimolecular reaction rate constants.}\label{tabapp}
 \begin{tabular}{llll}
  \hline\hline
 & {\ce{CH4 + OH}} &   {\ce{CD4 + OH}} &  {\ce{CH4 + OD}} \\
 &   {$T_\text{C} = 323.5$ K } & {$T_\text{C} = 240.2 $ K } & {$T_\text{C} = 323.1$ K } \\
 \hline
$T$ & $k$ & $k$ & $k$ \\
(K) & (s$^{-1}$) & (s$^{-1}$) & (s$^{-1}$) \\
  \hline
 300 &1.35 $\times$ $10^{-13}$   &                              & 1.78 $\times$ $10^{-13}$                    \\
 240 &1.80 $\times$ $10^{-14}$   &                              & 2.54 $\times$ $10^{-14}$                  \\
 230 &                           & 1.14 $\times$ $10^{-15}$     &                               \\
 220 &                           & 6.46 $\times$ $10^{-16}$     &  1.22 $\times$ $10^{-14}$ \\
 200 & 3.92 $\times$ $10^{-15}$  & 1.91 $\times$ $10^{-16}$     & 5.89 $\times$ $10^{-15}$ \\
 190 &                           & 1.00 $\times$ $10^{-16}$     & 4.20 $\times$ $10^{-15}$ \\
 180 &                           &  5.15 $\times$ $10^{-17}$    & 3.14 $\times$ $10^{-15}$ \\
 170 &                           & 2.60 $\times$ $10^{-17}$     & \\
 160 &                           & 1.30 $\times$ $10^{-17}$     & \\
 150 &                           & 6.48 $\times$ $10^{-18}$     & \\
 140 & 2.37 $\times$ $10^{-16}$  &                              & 4.23 $\times$ $10^{-16}$ \\
 130 &1.07 $\times$ $10^{-16}$   & 1.94 $\times$ $10^{-18}$     & 1.90 $\times$ $10^{-16}$                  \\
 120 &5.16 $\times$ $10^{-17}$   &                              &  9.61 $\times$ $10^{-17}$          \\
 110 &2.64 $\times$ $10^{-17}$   & 3.72 $\times$ $10^{-19}$     & 4.96 $\times$ $10^{-17}$                  \\
 100 &1.36 $\times$ $10^{-17}$   & 1.00 $\times$ $10^{-19}$     & 2.80 $\times$ $10^{-17}$                  \\
 95  &                           & 5.68 $\times$ $10^{-20}$     & 2.14 $\times$ $10^{-17}$                  \\
 90  &8.68 $\times$ $10^{-18}$   & 3.33 $\times$ $10^{-20}$     &  1.68 $\times$ $10^{-17}$                  \\                      
 85  &6.74 $\times$ $10^{-18}$   & 2.02 $\times$ $10^{-20}$     & 1.36 $\times$ $10^{-17}$                   \\                       
 80  &                           & 1.27 $\times$ $10^{-20}$     & \\
 75  &                           & 8.36 $\times$ $10^{-21}$     & \\
 70  &                           & 5.72 $\times$ $10^{-21}$     & \\
  \hline
  \end{tabular}
\end{table}
 
\end{appendix}
 
\end{document}